\documentclass[twocolumn,showpacs,aps,prl,superscriptaddress,floatfix]{revtex4}

\usepackage{epsfig}
\usepackage{graphics}
\usepackage{graphicx}
\usepackage{epic}
\usepackage{floatflt}
\usepackage{dcolumn}
\usepackage{amsmath}
\input babarsym.tex

\long\def\inst#1{\par\nobreak\kern 4pt\nobreak
    {\it #1}\par\vskip 10pt plus 3pt minus 3pt}

\def\sss{\scriptscriptstyle}
\def\barpd{{\raise.35ex\hbox{${\sss (}$}}--{\raise.35ex\hbox{${\sss )}$}}}
\def\dbarp{\hbox{$D^{0}$\kern-1.3em\raise1.5ex\hbox{\barpd}}}
\def\dbarpnozero{\hbox{$D$\kern-0.85em\raise1.5ex\hbox{\barpd}}}

\newcommand{\BaBarYear}      {06}
\newcommand{\BaBarNumber}    {003}
\newcommand{\SLACPubNumber}  {11807}

\def\de {\ensuremath{{\rm \Delta}E}\xspace}

\def\BTODK* {{\boldmath $B^-\to D^0K^{*-}$}} 

\begin{document}

\noindent
 \babar-PUB-\BaBarYear/\BaBarNumber \\
 SLAC-PUB-\SLACPubNumber \\

\title{Measurement of the \BTODK* ~branching fraction}
%
\author{B.~Aubert}
\author{R.~Barate}
\author{D.~Boutigny}
\author{F.~Couderc}
\author{Y.~Karyotakis}
\author{J.~P.~Lees}
\author{V.~Poireau}
\author{V.~Tisserand}
\author{A.~Zghiche}
\affiliation{Laboratoire de Physique des Particules, F-74941 Annecy-le-Vieux, France }
\author{E.~Grauges}
\affiliation{Universitat de Barcelona, Facultat de Fisica Dept. ECM, E-08028 Barcelona, Spain }
% ...
\collaboration{The \babar\ Collaboration}
\noaffiliation

\vspace{5mm}

\begin{abstract}

From a sample of 232 million $\FourS\to\BB$ events collected with the
\babar\ detector at the \pep2 \BF\ in 1999--2004,
we measure the $\Bm\to\Dz K^{*-}(892)$ decay branching fraction using
events where the $\Kstarm$ is reconstructed in the $\KS\pim$ mode and
the $\Dz$ in the $\Km\pip$, $\Km\pip\piz$, and $\Km\pip\pip\pim$ channels:
${\cal B}(\Bm\to\Dz K^{*-}(892)) = (\ 5.29 \pm 0.30\ ({\rm stat}) \pm 0.34\ ({\rm syst})\ ) \times 10^{-4}.$ 
\end{abstract}


\maketitle

The decays $\Bm\to\Dz K^{(*)-}$~\cite{chargeconj} 
are of interest because of their relevance
to the Cabibbo-Kobayashi-Maskawa (CKM) model~\cite{CKM} of quark-flavor mixing. 
Interference 
effects in specific $\Dz$ final states
offer a means of observing direct \CP violation
governed by the angle $\gamma=$arg($-V_{ud}V^*_{ub}/V_{cd}V^*_{cb}$)
~\cite{gamma1}, where $V$ is the CKM matrix. 
One way to access $\gamma$ is to compare the $B^-\to D^0K^{*-}$
branching fraction  to the \CP-averaged
branching fraction for a $B^-$ to decay into a $D^0K^{*-}$ where the $D^0$ decays into a
 \CP eigenstate~\cite{gamma2}. Thus a precise
determination of the $B^-\to D^0K^{*-}$ branching fraction provides the reference 
for direct \CP violation measurements.

The  decay $\Bm\to\Dz\Kstarm$ was first observed by
CLEO~\cite{cleo_bf}, and later  by \babar~\cite{babar_bf}. 
In this paper
we present a new measurement of 
the branching fraction ${\cal B}(\Bm\to\Dz\Kstarm)$  obtained with 2.7 times
more data than used for the previous \babar\ measurement.\par

This analysis uses data collected with the \babar\ detector at the \pep2\ e$^+$e$^-$ storage ring. 
 The data corresponds to an integrated luminosity of 211 \invfb at the \FourS
peak (232 million \BB\ pairs) and 16 \invfb\ at center-of-mass energy 40~\mev below the resonance.
\par

The \babar\ detector is described in detail in~\cite{babar}.
We give here a brief description of the components relevant to this analysis.
Charged-particle trajectories are measured by a five-layer 
double-sided silicon vertex tracker (SVT) and a 
40-layer drift chamber (DCH) inside a 1.5~T solenoid.
Charged-particle identification is achieved by combining measurements of the light
detected  in a ring-imaging Cherenkov
device (DIRC) with measurements of the ionization energy loss (\dedx ) measured in the DCH and SVT.
Photons are detected in a CsI(Tl) electromagnetic calorimeter (EMC) inside the coil.
We use GEANT4~\cite{geant4} based software to simulate the detector
response and account for the varying beam and environmental conditions. 

To reconstruct $\Bm\to\Dz\Kstarm$ decays we select \Kstarm\ candidates
in the $\Kstarm\to\KS\pim$ mode and \Dz candidates in three decay
channels: $\Dz\to\Km\pip$, $\Km\pip\piz$, and $\Km\pip\pip\pim$.  
Our event selection follows closely the one reported in~\cite{babar_glw}. \KS candidates are formed from oppositely
charged tracks assumed to be pions with a reconstructed invariant mass
within 13~\mevcc\ (four standard deviations) of the known \KS
mass, $m_{\KS}$~\cite{pdg2004}. The \KS candidates are fitted so 
that their invariant mass equals $m_{\KS}$ (mass constraint).
We further require their flight direction and distance to be consistent with a \KS\ coming from the interaction point.
The \KS\ candidate's flight path and momentum vectors must make an acute angle
and the flight length in the plane transverse to the beam must be at least three times larger than its uncertainty. 
\Kstarm candidates are formed from a \KS and a charged particle, which are required to originate from a common vertex.
We select \Kstarm candidates which have an invariant mass within 75~\mevcc of the known value~\cite{pdg2004}.
Finally, since the \Kstarm in $\Bm\to\Dz\Kstarm$ is polarized, we require the helicity
angle $\theta_H$ to satisfy $|$cos~$\theta_{H}| \geq 0.35$, where
$\theta_{H}$ is the angle in the \Kstarm\ rest frame between the daughter pion and the parent \B momentum. 
The helicity distribution discriminates well between a \B meson decay and an event from the $\epem \to \qqbar\ (q\in \{u,d,s,c\})$ 
continuum, since the former is distributed as $\cos^2\theta_{H}$ and the latter is almost flat.\par
In order to reconstruct the $\piz$ of the $\Dz\to\Km\pip\piz$ 
channel, we combine pairs of photons to form candidates with a total energy
greater than 200~\mev and an invariant mass between 
125 and 145~\mevcc. A mass-constrained fit is applied to the selected
\piz\ candidates. 
All \Dz candidates are mass- and vertex-constrained. Particle identification is required for the charged kaons. 
We select \Dz\ candidates with an
unconstrained invariant mass, $m_{\Dz}$, differing from the world average mass, $m^{PDG}_{\Dz}$, by less than 12~\mevcc 
for all channels except $\Km\pip\piz$ where we require $-29<m_{\Dz}-m^{PDG}_{\Dz}<+24~\mevcc$.
To reduce combinatorial background in this channel, we further select candidates in the regions of
the Dalitz plane enhanced by the $K^{*-}(892)$, $K^{*0}(892)$ and $\rho^+(770)$
resonances using amplitudes and phases measured by the CLEO experiment~\cite{DalitzWeight}.
In order to reduce the background from random two track combinations that
have masses consistent with a $D^0$ we also require,  for the $\Dz\to \Km\pip$ channel, $|$cos~$ \theta_{D}| \leq 0.9$, where
$\theta_{D}$ is the angle in the \Dz rest frame between the daughter kaon and the parent \B momentum. 
Finally, we perform a geometric fit on the \B candidate which constrains the \Dz , the \KS , and the charged pion from the \Kstarm
to originate from a single vertex.

To suppress continuum background we require $|$cos~$ \theta^*_B| \leq 0.9$, where
$\theta^*_B$ is defined as the angle between the \B candidate momentum in the \FourS\ rest frame and the beam axis.  
The distribution in $\cos\theta^*_B$ is flat for $q\bar{q}$ events, while for \B mesons it 
follows a $\sin^2\theta^*_B$ distribution.
We also use global event shape variables
to distinguish between \qqbar\ continuum events which have a two-jet topology in the \FourS\ rest frame and \BB\ events which
are more spherical. We require $|$cos~$ \theta^*_{T}| \leq 0.9$ where $\theta^*_{T}$ is
the angle between the thrust axes of the \B candidate and that of the rest of the event.
We construct a linear (Fisher) discriminant~\cite{fisher} from cos~$\theta^*_{T}$ 
and the $L_0$, $L_2$ monomials (see below) describing the energy flow in the rest of the event, as in~\cite{muriel}. 
In the center-of-mass frame (CM) we define $L_j=\Sigma_i p^*_i |\cos \theta^*_i|^j$, where $i$ indexes the charged and 
neutral particles in the event once those from the \B  candidate are removed, and $\theta^*_i$ 
is the angle of the CM-momentum $p^*_i$ with the thrust axis of the \B meson candidate.

We identify \B candidates using two nearly independent kinematic variables: the beam-energy-substituted mass
$\mes=\sqrt{(s/2+{\bf p_0 \cdot p_B})^2/E_0^2-p_B^2}$ and the energy difference $\Delta E=E_B^*-\sqrt{s}/2$, 
where $E$ and $p$ are energy and momentum, the subscripts 0 and $B$
refer to the \epem-beam-system and the \B candidate in the lab frame, respectively; 
$s$ is the square of the CM energy, and the asterisk labels the CM frame. 

In those events where we find more than one acceptable \B candidate 
(less than $25$\% of selected events depending on the \Dz mode), 
we choose the one with the smallest $\chi^2$ formed from the differences of the measured 
and world average \Dz and \Kstarm masses 
scaled by the mass resolution which includes the experimental resolution and, for the \Kstarm, its natural width. 
Simulations show that no bias is introduced by this choice and the correct candidate is picked at least 80\% of the time.
According to simulation of signal events, the total reconstruction efficiencies  are:
13.3\%, 4.6\%, and 9.0\% for the $\Dz\to\Km\pip$, $\Km\pip\piz$, and $\Km\pip\pip\pim$ modes, respectively.

To study \BB\ backgrounds we look at sideband regions away from the signal region in \de and  $m_{\Dz}$. 
The \de distributions are centered around zero for signal with a resolution between 11 and 13~\mev\ for all three channels. 
We define a signal region $|\Delta E| < 25$~\mev.
We also define a \de\ sideband in the  intervals $-100 \leq \de \leq -60 \mev$ and $60 \leq \de \leq 200 \mev $.  
The lower limit ($-100$~\mev) is chosen to avoid selecting a region of high background coming from $B^-\to\Dstar\Kstarm$.
In this \de\ sideband we see no significant evidence of a background peaking near the $B$ mass in $m_{ES}$ 
which could leak into the signal region.
The sideband region in $m_{\Dz}$ is defined by requiring that this quantity differs from the $\Dz$ mass peak by more than four standard deviations.
It provides sensitivity to doubly-peaking background sources that mimic signal both in \de\ and \mes. This pollution comes from
either charmed or charmless \B meson decays that do not contain a true \Dz. Since many of the possible contributions
to this background are not well known, we attempt to measure its size by including the $m_{\Dz}$ sideband in the fit described below.\par

An unbinned extended maximum likelihood fit to \mes distributions in the range $5.2\leq\mes\leq 5.3$~\gevcc 
is used to determine the event yields.
For signal modes, the \mes\ distributions are described by a
Gaussian function $\mathcal{G}$ centered at the \B mass with resolution ($\sigma$), averaged over the three $D^0$ decay modes,
 of 2.7 MeV/c$^2$.
For each $D^0$ decay  mode $k$ (=1, 2, 3) we determine the mean and sigma of the Gaussian $\mathcal{G}_k$ by fitting to the data. The
combinatorial background in the \mes\ distribution is modeled with a threshold function $\mathcal{A}_k$~\cite{argus}.
Its shape is governed by one parameter $\xi_k$ that is
left free in the fit for each $D^0$ decay mode.
We fit simultaneously \mes distributions of nine samples: the $\Km\pip$, the $\Km\pip\piz$ and $\Km\pip\pim\pip$
samples for ({\it i}) the $\Delta E$ signal region, ({\it ii}) the $m_{\Dz}$ sideband and ({\it iii}) the \de sideband.
We fit three probability density functions (PDF) weighted by the unknown event yields. 
For the \DeltaE sideband, we use $\mathcal{A}_k$. For the $m_{\Dz}$ sideband we use  $ N^k_{{\rm noP}} \cdot
\mathcal{A}_k$ + $N^k_{{\rm DP}} \cdot \mathcal{G}_k$, where $\mathcal{G}_k$ accounts for the doubly-peaking \B decays. 
For the signal region PDF we use $N^k_{q\bar{q}} \cdot \mathcal{A}_k+ \kappa N^k_{{\rm DP}} \cdot
\mathcal{G}_k+ N^k_{{\rm sig}} \cdot \mathcal{G}_k$, where  $\kappa$
is the ratio of the $m_{\Dz}$ signal-window to sideband
widths and $N^k_{{\rm sig}}$ is the number of $\Bm\to\Dz\Kstarm$ signal events. 
The \DeltaE sideband sample helps define the shape of the background function $\mathcal{A}_k$. 
We assume that the \B decays found in the $m_{\Dz}$ sideband have the same final states as the signal so
we use the same Gaussian shape for the doubly-peaking \B background.

\begin{figure}[t!]\begin{center}
\setlength{\unitlength}{1mm}
\begin{picture}(80,90)(0,0)
\jput(0,0){\epsfig{file=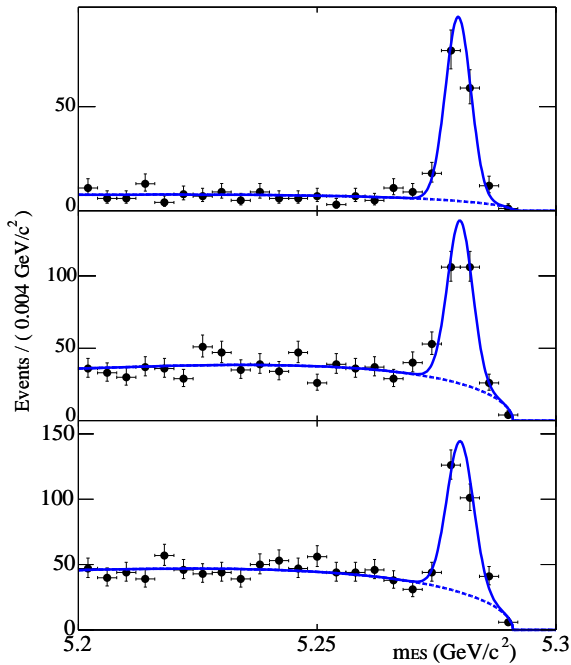,width=85mm}}
\end{picture}
\caption{
Distributions of \mes in the signal region for $\Bm \to \Dz \Kstarm$
decays where $\Dz \to \Km\pip$ (top), $\Km\pip\piz$ (middle), and
$\Km\pip\pim\pip$ (bottom). 
The dashed curve indicates the contribution from the combinatorial background and the peaking
\B-background which is estimated from a simultaneous fit to the \Dz sideband (not shown).
}\label{fig:bffit}\end{center}\end{figure}

\begin{table}[htbp]\begin{center}
\caption{\label{tab:nominalfitresult}
Results from the fit and quantities used to derive the $\Bm \to \Dz
\Kstarm$ branching fraction. For each channel we give the event yield
resulting from the fit, the efficiency, and the branching
fraction measurement, in units of 10$^{-4}$, derived using Eq.~\ref{eq:bf}. 
The uncertainties are statistical only.
}
\begin{tabular}{lccc}
\hline\hline
\multicolumn{1}{c}{}      & $\Km\pip$     &  $\Km\pip\piz$  & $\Km\pip\pim\pip$  \\
\hline

Yield     & 144 $\pm$ 13   & 185 $\pm$ 19     & 195 $\pm$ 18 \\
Efficiency& 13.30\% & 4.60\% & 8.99\% \\
${\cal B}(\Bm\to\Dz\Kstarm)$ & 5.15$\pm$0.47 & 5.65$\pm$0.57 & 5.24$\pm$0.49 \\

\hline
\end{tabular}
\end{center}\end{table}

The fit results are shown graphically in Fig.~\ref{fig:bffit} and
numerically in Table~\ref{tab:nominalfitresult}.
For each channel $k$, a measurement $\BR_k$ of the branching fraction 
$\BR(\Bm \to \Dz \Kstarm)$ is derived as follows:
\begin{equation}
\label{eq:bf}
\BR_{k} =\frac{N(\Dz\to X_k)\cdot f} {N_{\Bpm}\cdot \varepsilon_k \cdot \BR_{\Kstarm}\cdot \BR(\Dz\to X_ k)},
\end{equation}
\noindent
where 
$N(\Dz\to X_ k)$ is the event yield from the fit, $f$ the fraction of $K^{*-}$'s in the sample
(discussed below), 
$N_{\Bpm}$ is the number of  charged \B mesons in the data sample, 
$\varepsilon_{k}$ is the efficiency to reconstruct $\Bm\to\Dz\Kstarm$ when
$\Dz\to X_k$,
$\BR_{\Kstarm}\equiv\BR(\Kstarm\to{\KS}\pi)\cdot \BR(\KS\to\pip\pim)$ and $\BR(\Dz\to X_k)$
 are the branching fractions of the \Kstarm and the \Dz.
We have assumed equal production of pairs of neutral and charged \B
mesons in \FourS decay.\par

Systematic effects arise from the difference between the actual
detector response for the data and the simulation model for the Monte Carlo.
Here the main effects stem from the  modeling of the tracking efficiency (1.2-1.3\% per track),
the \KS\ reconstruction efficiency (2\% per \KS), the \piz reconstruction efficiency for
the $\Km\pip\piz$ channel (3\%) and the efficiency and misidentification probabilities from the particle identification (2\% per kaon). 
A study of a high-statistics $\Bm\to\Dz\pim$ control sample shows
excellent agreement between the data and Monte Carlo sample except for the distributions of \DeltaE and the continuum-suppression
Fisher discriminant. For these variables, differences of up to $(2.5\pm1.1)\%$ are measured between the data and Monte Carlo.
Suitable corrections to the efficiencies are therefore applied and systematic errors assigned.
The $K^{*-}$ helicity angle distributions differ significantly between
data and simulation because of the non-resonant background under the
\Kstarm peak. We describe below how we subtract this background. 
For the pure \Kstarm events, we estimate that the residual discrepancy between data
and simulation in the helicity to be less than 1.6\%.
We determine using simulations that the \mes signal PDFs deviate from
the single Gaussian shape by less than 0.1\%. 
Substantial systematic uncertainties come from the measured \Dz
branching fractions~\cite{pdg2004} and the number of $B^{\pm}$ pairs in the sample.

\begin{figure}[t!]\begin{center}
\setlength{\unitlength}{1mm}
\begin{picture}(80,55)(0,0)
\jput(0,0){\epsfig{file=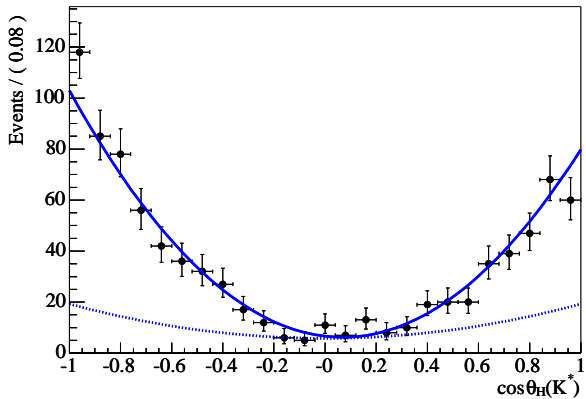,width=85mm}}
\end{picture}
\caption{
Acceptance corrected distribution of $\cos~\theta_{H}$(\Kstarm). 
The solid line is a fit to a model which includes P-wave and S-wave interference.
The dotted line shows the combinatorial background as estimated from the data.
}
\label{fig:helicity}\end{center}\end{figure}

\begin{figure}[t!]\begin{center}
\setlength{\unitlength}{1mm}
\begin{picture}(80,55)(0,0)
\jput(0,0){\epsfig{file=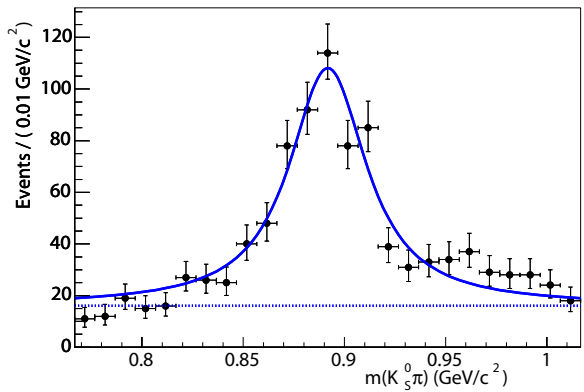,width=85mm}}
\end{picture}
\caption{
Invariant mass of $K^0_s\pi^-$ combinations with all other analysis cuts applied. 
The solid curve is a Breit-Wigner line shape including
detector resolution. The dotted line shows the combinatorial background.
}\label{fig:k*mass}\end{center}\end{figure}

 The observed number of signal events must be corrected for the  non-resonant $\KS \pi$ pairs under the
\Kstarm. When we remove the requirement on the \Kstarm helicity angle, we see that 
the \Kstarm helicity distribution (Fig.~\ref{fig:helicity}) of the selected events manifests a
forward-backward asymmetry that indicates an interference with a $\KS \pi$ background~\cite{babar_glw, swave}.
We model the $\KS\pi^-$ system with a P-wave and an S-wave component.
 The P-wave  
mass dependence
is described by a relativistic Breit-Wigner while the S-wave piece 
is assumed to be a complex constant.
This model is fitted to the data and shown in Fig.~\ref{fig:helicity} along 
with an estimate of the combinatorial background. 
Neglecting higher resonances, the number of $\KS\pim$ peaking
background events is $(4 \pm 1)\%$ of the total measured number of signal events.
We do not quote a systematic error on the contributions of the neglected partial waves 
(non-$K^*$ $P$-wave and higher order waves) since their expected rates in the $K\pi$ mass window are
far below that of the $S$-wave~\cite{swave}. In Fig.~\ref{fig:k*mass} we see that a
relativistic Breit-Wigner gives a fair description of the resonance
structure in the $\KS\pi^-$ mass spectrum ($\chi^2$=26.8 for 20 degrees of freedom).

\begin{table}[h!]\begin{center}
\begin{footnotesize}
\caption{\label{tab:bfsyst} 
Systematic uncertainties. $X_k$ refers to the $\Dz$ decay modes given in the 
columns. $\BR_{\Kstarm}$ is the branching fraction of the observed
$\Kstarm\to\KS\pim,\ \KS\to\pip\pim$ decay chain. 
}
\begin{tabular}{lccc}
\hline\hline
Source          &$\Km\pip$       &$\Km\pip\piz$     &$\Km\pip\pim\pip$\\
\cline{1-4}
Tracking efficiency        & 3.8\%       &    3.8\%       &   6.3\%\\
\piz efficiency            &  -          &    3.1\%       &    -  \\
Particle Identification    & 2.0\%       &    2.0\%       &   2.0\%\\
\KS efficiency             & 1.6\%       &    1.9\%       &   1.8\%\\
$\cos\theta_{H}(\Kstarm)$   & 1.6\%       &    1.6\%       &   1.6\%\\
Fisher                     & 1.1\%       &    1.1\%       &   1.1\%\\
\DeltaE                  & 1.9\%       &    1.8\%       &   2.0\%\\
\mes PDF shape             & 0.1\%       &    0.1\%       &   0.1\%\\
Number of $\Bpm$             & 1.1\%       &    1.1\%       &   1.1\%\\
Simulation statistics                    & 0.9\%       &    1.4\%       &   1.0\%\\
$\BR_{\Kstarm}$~\cite{pdg2004}           & 0.2\%       &    0.2\%       &   0.2\%\\
$\BR(\Dz\to X_k)$~\cite{pdg2004}           & 2.4\%       &    6.2\%       &   4.2\%\\
\KS \pim S-wave subtraction              & 1.1\%       &    1.1\%       &   1.1\%\\
\cline{1-4}
Total systematic error                   & 6.1\%       &    9.0\%       &   8.7\%\\
\hline 
\end{tabular}
\end{footnotesize}
\end{center}
\end{table}

All sources of systematic uncertainties are listed for each mode in~\tabref{bfsyst}.
With the exception of $\DeltaE$ and simulation statistics
the systematic error sources listed in~\tabref{bfsyst} are correlated among the
different $D^0$ modes. 
We use the procedure discussed in~\cite{avg} to form a weighted average of the three $D^0$ 
decay modes and determine:

\begin{equation}
{\cal B}(\Bm\to\Dz\Kstarm) = (\ 5.29 \pm 0.30 \pm 0.34 \ ) \times 10^{-4}.\nonumber
\end{equation}
The first error is statistical and the second is systematic.
We have compared the results from this analysis using the same data set
as in our previously published analysis~\cite{babar_bf}. The two analyses use different selection 
criteria and therefore find different numbers of events. The results from the
two analyses are consistent to within a half
of a (statistical) standard deviation.
We have also calculated the branching fraction for the two data sets obtained since the previous analysis. 
The measurement in each set is consistent with, although lower than the value obtained in~\cite{babar_bf}.
 This result supersedes our previously published result.

In summary, we have measured the branching fraction of the decay
$\Bm\to\Dz\Kstarm$ in the $\Dz\KS\pi^-$ final state and observed the interference of the
\Kstarm with a small non-resonant $\KS\pi$ background.

We are grateful for the excellent luminosity and machine conditions
provided by our \pep2\ colleagues, 
and for the substantial dedicated effort from
the computing organizations that support \babar.
The collaborating institutions wish to thank 
SLAC for its support and kind hospitality. 
This work is supported by
DOE
and NSF (USA),
NSERC (Canada),
IHEP (China),
CEA and
CNRS-IN2P3
(France),
BMBF and DFG
(Germany),
INFN (Italy),
FOM (The Netherlands),
NFR (Norway),
MIST (Russia), and
PPARC (United Kingdom). 
Individuals have received support from CONACyT (Mexico), 
Marie Curie EIF (European Union),
the A.~P.~Sloan Foundation, 
the Research Corporation,
and the Alexander von Humboldt Foundation.


\begin{thebibliography}{99}
\bibitem{chargeconj} Reference to a charge conjugate mode
 is implied throughout the paper unless otherwise stated.
\end{thebibliography}
\end{document}